\documentclass[10pt,conference,compsoc,onecolumn]{IEEEtran}

\usepackage{times,amsmath,epsfig}
\usepackage[english]{babel}
\usepackage{amsmath,amsfonts,amsthm} 
\usepackage[hang,small,labelfont=bf,up,textfont=it]{caption}
\usepackage{booktabs} 
\usepackage{lastpage}
\usepackage{graphicx} 

\usepackage{enumitem}
\setlist{noitemsep} 
\usepackage{pstricks}
\usepackage{amssymb}
\usepackage[utf8]{inputenc}
\usepackage{multirow}
\usepackage{amsmath}
\usepackage{mathtools}
\usepackage{tikz}
\usepackage{bclogo}
\usepackage{color, colortbl}
\usepackage{amsthm}
\usepackage{float}
\usepackage{color}
\usepackage[ruled,vlined,linesnumbered]{algorithm2e}
\usepackage{subcaption}
\usepackage{caption,setspace}
\usepackage[english]{babel} 

\newcommand{\uprove}{U-prove }
\newcommand{\idemix}{Idemix }
\newcommand{\itwopa}{I2PA }

\newcommand{\idemixs}{Idemix's }
\newcommand{\itwopas}{I2PA's }

\newcommand{\uprovev}{U-prove, }
\newcommand{\idemixv}{Idemix, }
\newcommand{\itwopav}{I2PA, }

\newcommand{\uprovep}{U-prove. }
\newcommand{\idemixp}{Idemix. }
\newcommand{\itwopap}{I2PA. }

\newcommand{\papertitle}{I2PA, U-prove, and Idemix: An Evaluation of Memory Usage and Computing Time Efficiency in an IoT Context}

\numberwithin{theorem}{section} 

\usepackage[utf8]{inputenc}
\usepackage{hyperref}
\usepackage{multirow}
\usepackage{graphicx}
\hypersetup{
	colorlinks=true,
	linkcolor=blue,
	filecolor=magenta,
	urlcolor=blue,
	citecolor=blue
}

\usepackage[ruled,vlined,linesnumbered]{algorithm2e}

\usepackage{geometry}
\usepackage[T1]{fontenc} 
\usepackage[utf8]{inputenc}

\usepackage{fancyhdr} 
\title{\papertitle}
\author{
{SENE Ibou{\small $~^{\#*1}$}, CISS Abdoul Aziz{\small $~^{\#2}$} and NIANG Oumar{\small $~^{\#3}$} }
\vspace{1.6mm}\\
\fontsize{10}{10}\selectfont\itshape
$^{\#}$\,Ecole Polytechnique de Thiès (E.P.T.), \\Laboratoire de Traitement de l’Information et des Systèmes Intelligents (LTISI),\\
PO Box A10, Thiès, Sénégal\\
\vspace{1.2mm}
\fontsize{10}{10}\selectfont\itshape
$^{*}$\,Université de Thiès (U.T.),\\ Ecole Doctorale Développement Durable et Société (ED2DS),\\
PO Box 967, Thiès, Sénégal\\
\fontsize{9}{9}\selectfont\ttfamily\upshape
$^{1}$\,senei@ept.sn,
$^{2}$\,aaciss@ept.sn,
$^{3}$\,oniang@ept.sn
}

\begin{document}
\maketitle

\begin{abstract}
The Internet of Things (IoT), in spite of its innumerable advantages, brings many challenges namely issues about users' privacy preservation and constraints about lightweight cryptography. Lightweight cryptography is of capital importance since IoT devices are qualified to be resource-constrained. To address these challenges, several Attribute-Based Credentials (ABC) schemes have been designed including \itwopav \uprovev and \idemixp Even though these schemes have very strong cryptographic bases, their performance in resource-constrained devices is a question that deserves special attention. Therefor, this paper aims to conduct a performance evaluation of these schemes on issuance and verification protocols regarding memory usage and computing time. Recorded results show that both \itwopa and \uprove present very interesting results regarding memory usage and computing time while \idemix presents very low performance  with regard to computing time compared to  \itwopa and \uprovep
\end{abstract}


%-----------------------
%INTRODUCTION
%-----------------------
\section{Introduction}
\label{label-introduction}

Out of several emerging technologies and concepts, the Internet of Things is a new paradigm that brings both challenges and opportunities \cite{chenyenkuang2012}. According to Ashton Kevin, to whom we owe the term "Internet of Things", the IoT has the potential to change the world, as did the Internet, maybe even more \cite{kevin-ashton-2009-iot}. The Internet of Things represents a vision in which the Internet extends into the real world embracing everyday objects \cite{mattern2010internet}. However, as mentioned above, it brings many challenges including issues about users' privacy preservation and constraints about lightweight cryptography \cite{isenei2paiotmdpijuin2019}. We are among those who think that the protection of privacy is a fundamental right and its loss would lead to the restriction of freedom \cite{toumia2018pretopologie}. Lightweight cryptography is a strong constraint because IoT devices are qualified to be resource-constrained. Indeed, these devices have three major constraints namely low energy autonomy, very limited storage capacity and very low computing power \cite{isenei2paiotmdpijuin2019}. From there, were designed several schemes and the most promising include \itwopa \cite{isenei2paiotmdpijuin2019}, \idemix \cite{camenisch2002design}, and \uprove \cite{paquin2011u}. These schemes are based on recognized robust cryptosystems. However, the question of their applicability in an IoT context, therefore in resource-constrained devices, is of capital importance. Roughly results recorded in \cite{isenei2paiotmdpijuin2019} on issuance and verification of credentials made up of 10 attributes show that \itwopa and \uprove are more efficient than \idemix  regarding computing time efficiency. However, what about memory usage and computing time efficiency on different number of attributes ? In this paper, we provide a deeper evaluation by regarding memory usage and computing time while issuing and verifying  credentials made up of 1, 5, and 10   attributes respectively.

The rest of this paper is organized as follows. Section \ref{label-background} is related to background review while related works are presented in Section \ref{label-related-work}. Section \ref{label-experimental-environment} describes experimental set-up whereas recorded results and discussions are depicted in Section \ref{label-result-discussion}. This paper is ended by a conclusion and perspectives in Section \ref{label-conclusion-future-works}.

%-----------------------
%BACKGROUND
%-----------------------
\section{Background Review}
\label{label-background}

We now recall few notions about Attribute-Based Credentials (ABC), Elliptic Curves Cryptography (ECC), Binary Scalar Multiplication (BSM), and Extended Homogeneous Coordinates (EHC). We refer readers to \cite{isenei2paiotmdpijuin2019, alpar2013credential, alpar2015attribute, aziz2015trends, josefsson2017edwards} for more details about discussed concepts in this section.

\subsection{Attribute-Based Credential}
\label{label-abc}

Attribute-Based Credentials are mechanisms of authentication that allow to flexibly and selectively authenticate different attributes about an entity without revealing additional information about that entity. As a result, they do not necessarily identify the user, as they only provide authentic assertions about the user \cite{isenei2paiotmdpijuin2019, alpar2013credential, alpar2015attribute}. They are building blocks that aim at protecting users' privacy preservation.

\subsection{Elliptic Curve Cryptography}
\label{label-ecc}

Elliptic Curve Cryptography (ECC) was presented independently by Koblitz \cite{koblitz1987elliptic} and Miller \cite{miller1985use} in the 1980s. Their structure of group and performance in computing time they offer make them a new direction in cryptography \cite{isenei2paiotmdpijuin2019}. They offer good level of security with smaller key size. They are also less Central Processing Unit (CPU) intensive so they are ideal for resource-constrained devices.

\subsection{Binary Scalar Multiplication}
\label{label-double-and-add}

 The fundamental operation of ECC is point scalar computation (also known as scalar multiplication) of the form \cite{isenei2paiotmdpijuin2019}: 
\[
Q=k.P=\underbrace{P+P+\ldots +P}_{\text{k times}}
\]

Security in ECC is based on Elliptic Curve Discrete Logarithm Problem (ECDLP) \cite{isenei2paiotmdpijuin2019, aziz2015trends} that can be summarized as follows. Given an elliptic curve $E$ defined over a finite field $\mathbb{F}_p$. Let $P, Q \in E (\mathbb{F}_p)$, find $k \in \mathbb{F}_q$, if it exists, such that $Q= k.P$ ($q$ denotes the order of $P$). Scalar multiplication can be performed efficiently when tackling small numbers. However, when numbers hold in many bits (160 for instance), this might take lot of time. Several methods have been designed so far to speed up these operations including the double-and-add algorithm also known as binary algorithm. This algorithm is a very elegant technique to perform multiplication of big numbers. Two versions of this algorithm exist that either scan the scalar in a left to right or right to left direction \cite{rivain2011fast}. Let $k$ be an integer such that $k_{ (10)}= ({k_{n} k_{n-1} \dots k_{1} k_{0}})_{ (2)},$ where $ k_{i }\in \{0,1\}, k_{n} = 1~and~ n \ge 1 $. The left to right version is described in "Algorithm \ref{algo_left_right}".

 \begin{algorithm}[h]
\SetAlgoLined 
\KwIn{ $P \in E (\mathbb{F}_{p}), k \in \mathbb{F}_{q} $ }
\KwResult{ $k.P \in E (\mathbb{F}_{p})$}
 R $\leftarrow $P\\
 \For{$i\leftarrow n-1$ \KwTo $0$}{
 R $\leftarrow $2.R\\
 \If{ $k_{i}=1$}{
 R $\leftarrow $R+P
 }
 }
\Return $R$
\caption{Left to right double-and-add}
\label{algo_left_right}
\end{algorithm}

The "Algorithm \ref{algo_left_right}" is simple, efficient and has an average complexity of $nD+\frac{n}{2}A$ (D and A denote respectively the number of double and add operations). Implementations of \itwopa and \uprove are based on this technique seeing its simplicity, its low complexity, and its easy implementation in place of other methods like Non-Adjacent Form (NAF) also known as Signed Binary Representation (SBR) which presents a more interesting complexity ($nD+\frac{n}{3}A$) but requires a supplementary treatment on the representation of the scalar. Let us consider a device with a processor  clocked at 1GHz and $k=2^{40}$. Computing $k.P$ with decimal representation of $k$ would require around 19 minutes while with binary representation, this would take less than 1 millisecond. We remind that the number of bits required to represent a positive integer $n$ in radix $2$ is at most equal to $ceil (log_{2} (n))+1$, where $ceil (x)$ denotes the rounds of $x$ up to the nearest integer. These results show how relevant it is to use this technique instead of decimal representation.

\subsection{Extended Homogeneous Coordinates}
\label{label-extended-homogeneous-coordinates}

According to our experimental parameters (see Section \ref{label-params}), Simon et al. \cite{josefsson2017edwards} recommended to use extended homogeneous coordinates (EHCs). In the EHCs representation, $ (x,y)$ is represented as $ (X:Y:Z:T)$ where $x=\frac{X}{Z}$, $y=\frac{Y}{Z}$ and $xy=\frac{T}{Z}$. The neutral point $ (0,1)$ is equivalent to $ (0:Z:Z:0)$ for any nonzero $Z$. Coordinates $ (X:Y:Z:T)$ and $ (\lambda X:\lambda Y:\lambda Z:\lambda T)$ are equivalent for any nonzero $\lambda$. EHMs avoid inversion operations and, as a result, improve computing time efficiency. We refer readers to \cite{josefsson2017edwards} for more details. Details of add and double formulas are presented respectively in "Algorithm \ref{algo_projective_cordinate_adding}" and "Algorithm \ref{algo_projective_cordinate_doubling}". 

\begin{figure}[h]
\centering
\begin{minipage}[b]{0.488\linewidth}
\begin{algorithm}[H]
\SetAlgoLined
\KwIn{ $P_1, P_2 \in E (\mathbb{F}_{p})$ }
\KwResult{ $P_1+P_2 \in E (\mathbb{F}_{p}) $}
 A $\leftarrow $ $ (Y_1-X_1) (Y_2-X_2)$ \\
 B $\leftarrow $ $ (Y_1+X_1) (Y_2+X_2)$ \\
 C $\leftarrow $ $2dT_1T_2$ \\
 D $\leftarrow $ $2Z_1Z_2$ \\
 E $\leftarrow $ $B-A$\\
 F $\leftarrow $ $D-C$\\
 G $\leftarrow $ $D+C$\\
 H $\leftarrow $ $B+A$\\
 $(X,Y,Z, T)\leftarrow (EF,GH,FG,EH)$
 \Return $ (X:Y:Z:T)$
 \caption{Add formula}
 \label{algo_projective_cordinate_adding}
\end{algorithm}
\end{minipage}
\hspace{0.1cm}
\begin{minipage}[b]{0.488\linewidth}
\begin{algorithm}[H]
\SetAlgoLined
\KwIn{ $P \in E (\mathbb{F}_{p})$ }
\KwResult{ $2.P\in E (\mathbb{F}_{p}) $}
 A $\leftarrow $ $X^2$ \\
 B $\leftarrow $ $Y^2$ \\
 C $\leftarrow $ $2Z^2$ \\
 D $\leftarrow $ $ (X+Y)^2$\\
 H $\leftarrow $ $B+A$\\
 E $\leftarrow $ $H-D$\\
 G $\leftarrow $ $A-B$\\
 F $\leftarrow $ $C+G$\\
 $(X^\prime,Y^\prime,Z^\prime, T^\prime)\leftarrow (EF,GH,FG,EH)$
 \Return $ (X^\prime:Y^\prime:Z^\prime:T^\prime)$
 \caption{Double formula}
 \label{algo_projective_cordinate_doubling}
\end{algorithm}
\end{minipage}
\end{figure}

%-----------------------------
%RELATED WORKS
%-----------------------------
\section{Related Works}
\label{label-related-work}

The Internet of Things brings both challenges and opportunities \cite{chenyenkuang2012}. Indeed, in an IoT context, performance, privacy preservation, and lightweight cryptography are key aspects that must be taken into account with special attention. To the best of our knowledge, the best way of protecting users' privacy preservation remains using Attributes-Based Credentials (ABC) also known as Privacy-ABC. Many ABC schemes have been designed so far including \itwopav \uprovev and \idemixp However, few works evaluate the efficiency of these schemes in an IoT context. On a theoretical level, authors of \cite{baldimtsi2013anonymous, camenisch2012efficient} have addressed the importance of computational efficiency in resource-constrained devices. Fatbardh et al. \cite{veseli2016evaluation} have evaluated the computational efficiency of \idemix and \uprovep Their results shown that \uprove is more efficient than \idemix for the User operation (proving) and in general when a credential has more attributes. They have also stated that \idemix is more efficient in the rest of the cases, especially when advanced presentation features are used. Their simulation uses a computer with a processor of 1.8 GHz Intel Core i7 and  both schemes are instantiated using the RSA cryptosystem. Veseli et al. \cite{vfso2015} have addressed storage and communication efficiency of \idemix and \uprovep Their results suggest that for storage, \idemix is more efficient than \uprovev since a single credential provides multiple-show unlinkability. They have also pointed out that, in terms of communication efficiency, \idemix is more efficient for issuance, whereas \uprove is more efficient for presentation of credentials. They have developed a number of experiments in Java, which have been executed on a computer with a processor of 1.8 GHz Intel Core i7 and schemes are based on the RSA cryptosystem. Vullers et al. \cite{vullers2013efficient} have presented an efficient selective disclosure on smart cards using \idemix (using the MULTOS platform). Their implementation is based on a 1024 bits security level. They asserted that \idemixs selective disclosure can be efficiently implemented on a smart card. Mostowski et al. \cite{mwvp2012lnicst} provided an efficient \uprove implementation for Anonymous Credentials on smart cards (Using the MULTOS platform). Their implementation aims at making the smart card independent of any other resources, either computational or storage.  Their performance results strongly support their idea to use a stand-alone \uprove smart card rather than the Microsoft device-protection approach, which seems to overlook the current capabilities of smart cards. SENE et al. \cite{isenei2paiotmdpijuin2019} have conducted a comparison of \itwopav \uprovev and \idemix on issuance and verification regarding computing time for credentials made up of 10 attributes. They have instantiated \uprove using ECC and their results have shown that \uprove presented more interesting results than \idemix regarding computing time on issuance and verification protocols. \itwopa and \uprove present nearby performance even though \itwopas results are more interesting. Although these works have presented interesting results, most of them have focused on the efficiency of a particular implementation of a particular technology, and on a particular platform. Some of them were interested in many schemes or many aspects of privacy preservation but used computer which does not give any idea on low-resource devices efficiency. To the best of our knowledge, this is the first contribution that evaluates efficiency of \itwopav \uprove and \idemix in an IoT context regarding computing time and memory usage. Furthermore, as far as we know, it is also the first one that considers ECC-based \uprove instantiation in low-resource devices.

%-----------------------
%EXPERIMENTAL ENVIRONEMENT
%-----------------------
\section{Experimental Setup}
\label{label-experimental-environment}

This section describes both hardware and software setup. It also describes curve and parameters used to perform this evaluation.

\subsection{Hardware Setup}

\begin{figure}[!htb]
 \centering{
 \rotatebox{90}{
 \includegraphics[width=0.28\textwidth]{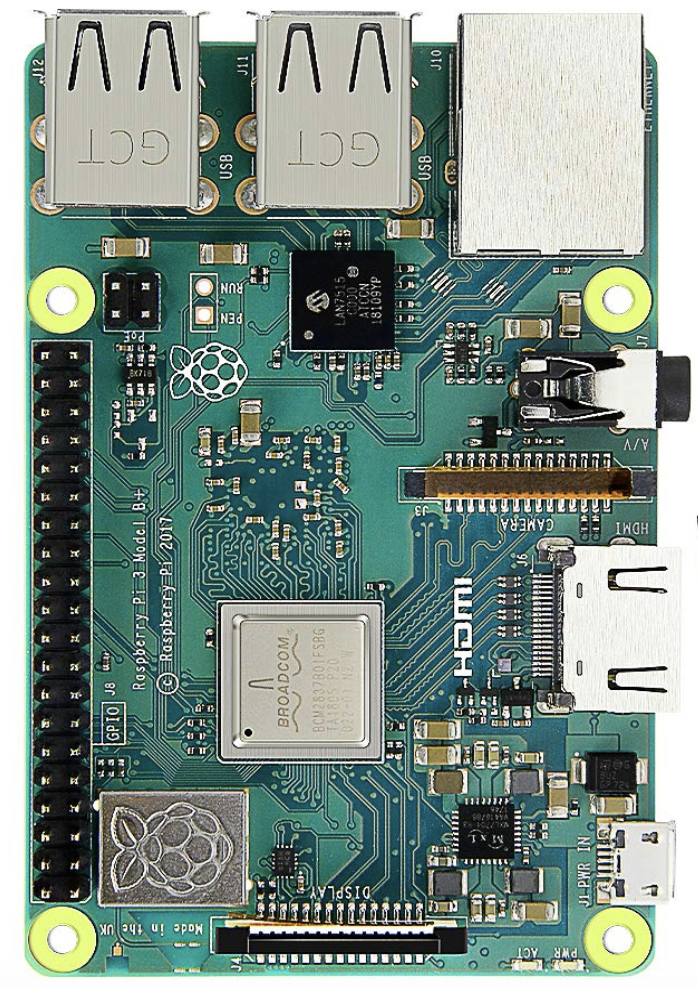}
 }
 \caption{Hardware environment}
 \label{fig:hardware-raspberry}
}
\end{figure}

The hardware setup consists of a smartphone and a Raspberry Pi. The Raspberry Pi ("Figure \ref{fig:hardware-raspberry}") is used to deploy both issuer and verifier. We describe some of its characteristics below: 

\begin{itemize}
\item { Model Pi 3 B+}
\item {1 Go of SDRAM LPDDR2}
\item{A 64-bit quad core processor clocked at 1.4 GHz}
\item {Raspbian operating system. }
\item {Dual Band 2.4 GHz and 5GHz IEEE 802.11. b/g/n/AC Wireless LAN}
\item {Enhanced Ethernet performance over USB 2.0 (maximum throughput of 300 Mbps)}
\end{itemize}

The smartphone ("Figure \ref{fig:screen-short-android-app}") acts as a user. Some of its characteristics are depicted below: 

\begin{itemize}
\item { Model TECNO SPARK KB7j}
\item { RAM 2GB}
\item { ROM 16GB}
\item {CPU 2.0GH*4 }
\item { Battery 3500 mAh}
\item { Memory 16GB}
\end{itemize}

\begin{figure}[!htb]
	\centering{
	\begin{subfigure}{0.27\textwidth}
	\centering{
		\includegraphics[width=2.8cm]{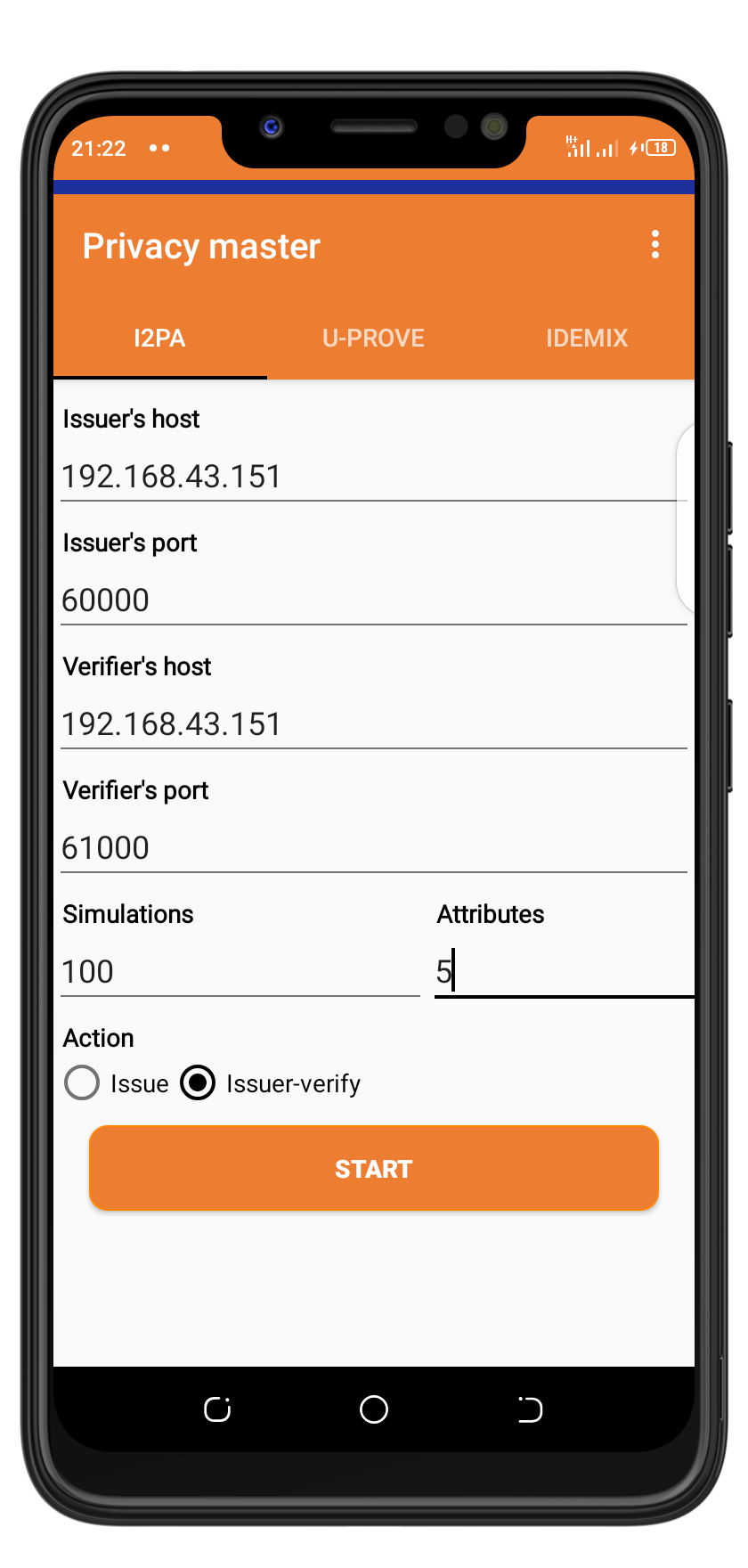}
		\caption{Initial view} 
	}
	\end{subfigure}
	\begin{subfigure}{0.27\textwidth} 
	\centering{
		\includegraphics[width=2.8cm]{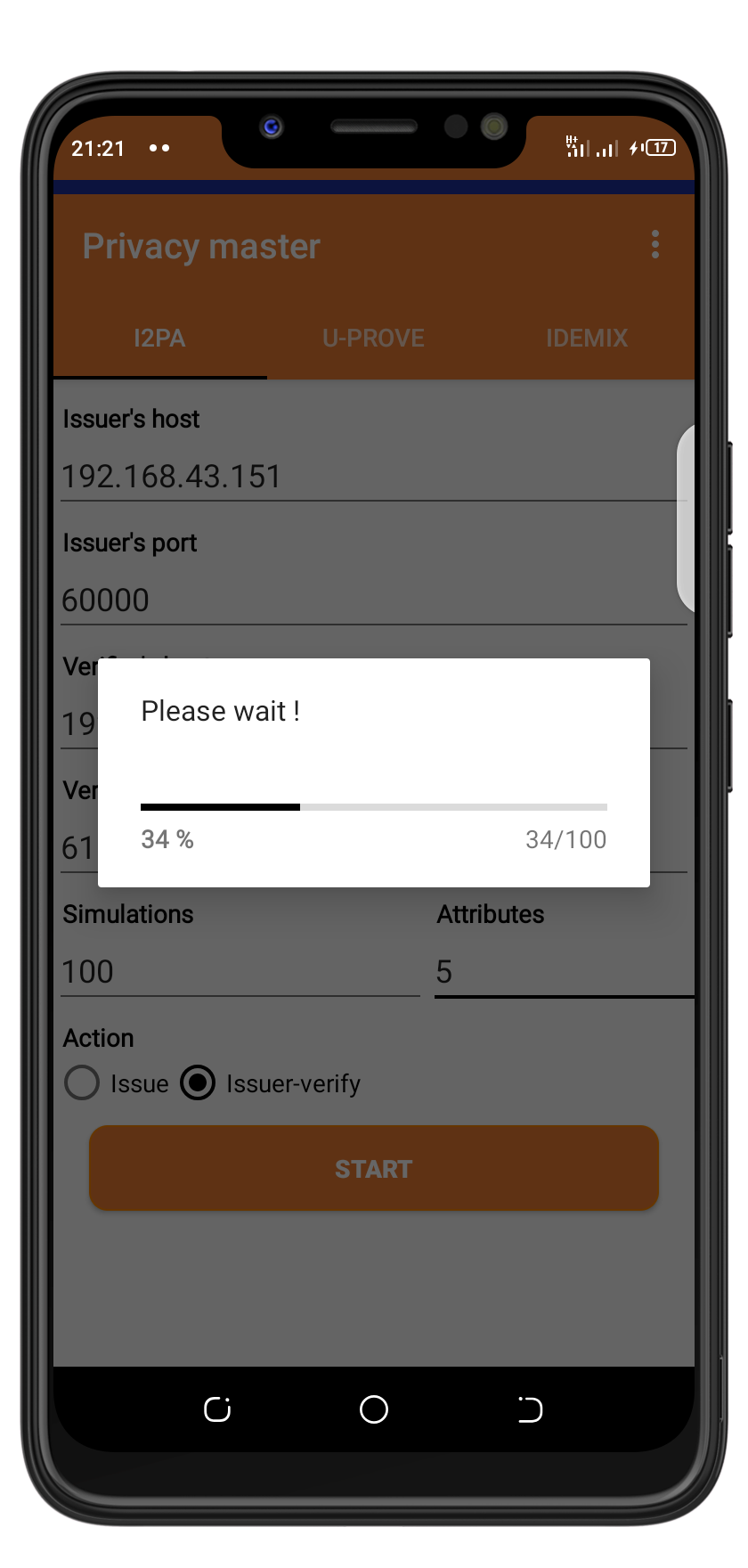}
		\caption{Processing view}
	}
	\end{subfigure}
	\vspace{1em} 
	\begin{subfigure}{0.27\textwidth} 
	\centering{
		\includegraphics[width=2.8cm]{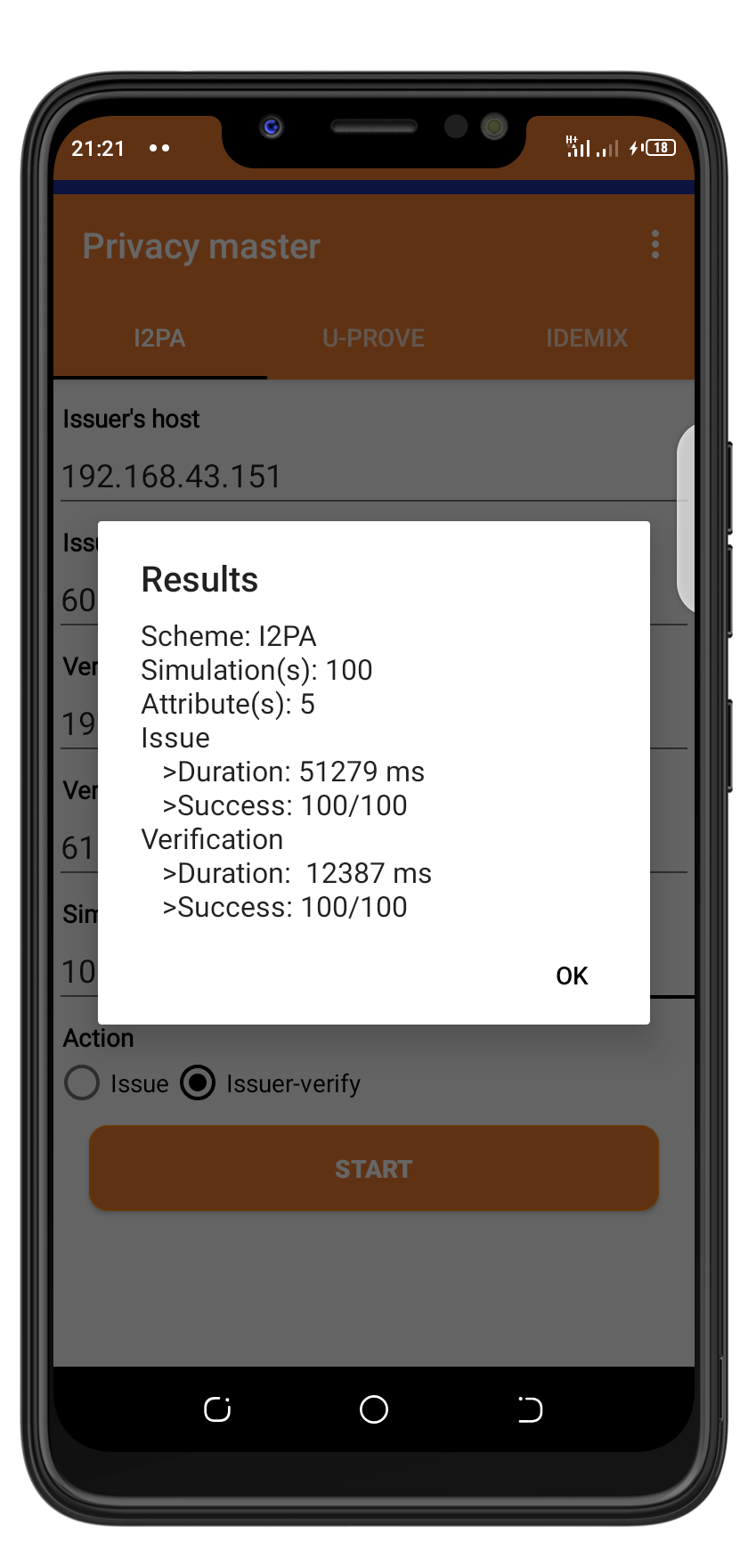}
		\caption{Result view} 
	}
	\end{subfigure}
	
	 \caption{Android application's screenshots.}
 \label{fig:screen-short-android-app}
 }
\end{figure}

\subsection{Software Setup}

\begin{figure}[!htb]
 \centering{
 \includegraphics[width=0.5\textwidth]{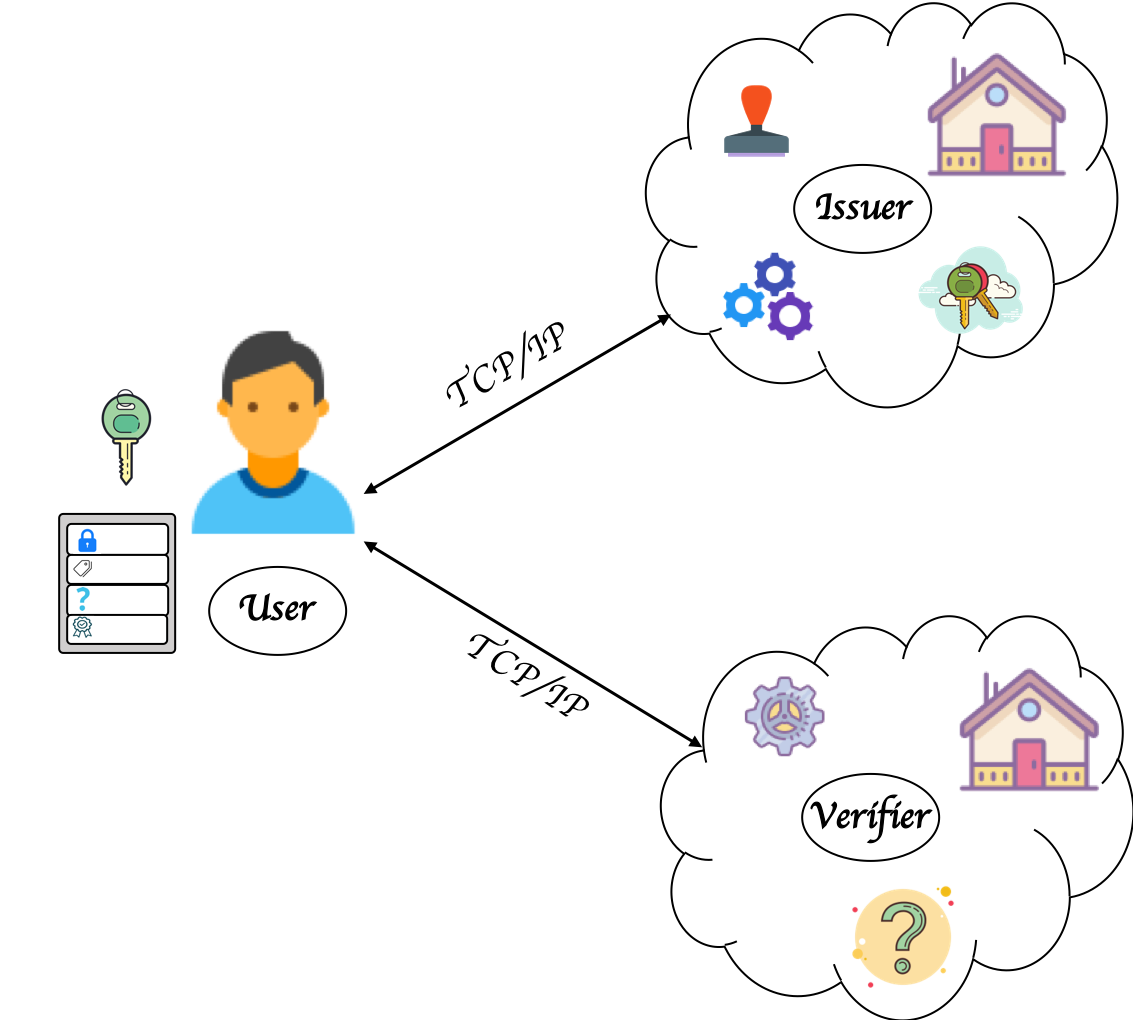}
 \caption{Software environment}
 \label{fig:i2pa-architecture-logicielle}
}
\end{figure}

The software environment is made up of three major components that are issuer, verifier, and user ("Figure \ref{fig:i2pa-architecture-logicielle}"). Issuer and verifier are Java Socket while user is an Android application. We are running both the issuer and the verifier on the same device (the Raspberry Pi) while the user is running on a smartphone. The "Figure \ref{fig:i2pa-architecture-logicielle}" is an overview of software components.

%-----------------------
%PARAMETER
%-----------------------
\subsection{Parameters}
\label{label-params}

Edwards' curves are known to offer better performances among all Elliptic Curve (EC) families \cite{liu2015performance}. The $Curve25519$ was introduced as an ECDH (Elliptic Curve Diffie-Hellman) function but it is known today as the underlying elliptic curve designed for use with ECDH key agreement scheme ($X25519$) or with ECDSA (Elliptic Curve Digital Signature Algorithm) signature ($Ed25519$). It was first introduced in its Montgomery form $E:  v^2 = u^3 + 486662u^{2} + u$ over the prime field defined by $p = 2^{255}-19$. This curve ensures a 128-bit security level as the fastest known attack on the discrete logarithm problem \cite{elhousni:hal-01942759}. Nowadays, it is used in Protocols, Networks, Operating Systems, Software, SSH Software, TLS Libraries, etc. \cite{curve-25519-usage}. Below, we describe parameters used in our performance evaluation and they are adapted from \cite{elhousni:hal-01942759}. The parameter $k$ defines keys' size for schemes \itwopa and \uprove while $k^{\prime}$ defines \idemixs keys size. The parameter $p$ defines the field $\mathbb{Z}_p$, $d$ defines the elliptic curve $E_d:x^2+y^2=1+dx^2y^2$. Values $x_0$ and $y_0$ define the coordinates of the base point $P$ with order $q$. The component $Z_0$ defines the third component in extended homogeneous coordinates of the base point. We refer reader to \cite{aziz2015trends, sinha2013performance} for keys' size justification. Parameters' values are depicted below:

\begin{itemize}
\item {$k = 160$}
\item {$k^{\prime} = 1024$}
\item {$p = 2^{255}-19$ }
\item {$d = 370957059346694393431380835087545651
89542113879843219016388785533085940283555$ }
\item {$x_0 = 15112221349535400772501151409588531
5114540126930418572060461132839498477622
02$}
\item {$y_0 = 46316835694926478169428394003475163
1413079938662562256157830336031652518559
60$}
\item {$Z_0 = 1$}
\item {$q=2^{252} + 2774231777737235353585193
 7790883648493$ }
\end{itemize}

At the core of ABC schemes, we have attributes. An attribute is a characteristic or a qualification of a person \cite{isenei2paiotmdpijuin2019}. It certifies that an entity has skill, knowledge, qualification, etc. An attribute certified by a third party is known as a claim. Whatever the nature of an attribute, it can be represented in a decimal format. Therefore, attributes' values used in this evaluation are described below:

\begin{itemize}
\item{$a_0$=3022871045856445402}
\item{$a_1$=2303921356947}
\item{$a_2$=63990592803}
\item{$a_3$=63188281798077}
\item{$a_4$=2334544185927680150715}
\item{$a_5$=72478959060716899515}
\item{$a_6$=132108418240270107954363}
\item{$a_7$=53359477949683103}
\item{$a_8$=393090009322226684739352798186683}
\item{$a_9$=2930303348526267}
\end{itemize}

%-----------------------
%RESULTS AND DISCUSSION
%-----------------------
\section{Results and Discussion}
\label{label-result-discussion}

This section depicts and comments results of our performance evaluation. Unless explicitly stated, time will always be expressed in milliseconds (ms) and memory in Megabyte (MB). 
It should also be noted that, for memory metrics,  all values are rounded to two decimal places. We remind that \uprove and \itwopa are instantiated using ECC as mentioned before. The implemented versions of \uprove and \idemix are based on schemes presented by Gergely Alp{\'a}r \cite{alpar2015attribute} while \itwopa implementation is based on the scheme presented by SENE et al. \cite{isenei2paiotmdpijuin2019}. We point out that every simulation is carried out with new random parameters except system's parameters and attributes' values.

\subsection{Limitations}
\label{label-limitations}

We note some limitations that should be taken into account while exploring results presented thereafter.

\begin{itemize}
\item {
Our results are based on the openly available versions of \uprove
and \idemixp
}
\item {
During the issuance phase and at user side, when the issuer takes lots of times to issue credential, recorded minima in terms of memory usage at user side may be biased. The user may remain idle for a while which considerably lowers used resources. This is the case with \idemix since its issuance requires lots of times (See "Figure \ref{fig:issue-time}").
}
\item {
During the verification phase, in order not to impact memory usage, we first issue a credential and then verify it immediately instead of storing all credentials that should be verified. This may  impact the recorded minima at verifier side if the issuance of a credential takes lots of times. The later may remain idle for a while what considerably lowers used resources. This is  the case with \idemix that requires a lot of times to issue a credential (See "Figure \ref{fig:issue-time}").
}
\end{itemize}

%-----------------------
%MEMORY USAGE
%-----------------------
\subsection{Memory Usage Evaluation}

This section describes results about memory usage. Figures presented below are recorded with VisualVM 1.3.9 \cite{VisualVM-2017} using "Tracer-Monitor Probes" plugin. Evaluations involve 100 simulations on issuance and verification of credentials made up of 1, 5, and 10 attributes respectively.

\subsubsection{Issuance}

In this section, we describe memory evaluation at issuer and user sides. At issuer side (respectively at user side), we evaluate the memory required to issue (respectively to get) a credential.

\begin{figure}[H]
 \centering{
 \includegraphics[width=0.9\linewidth]{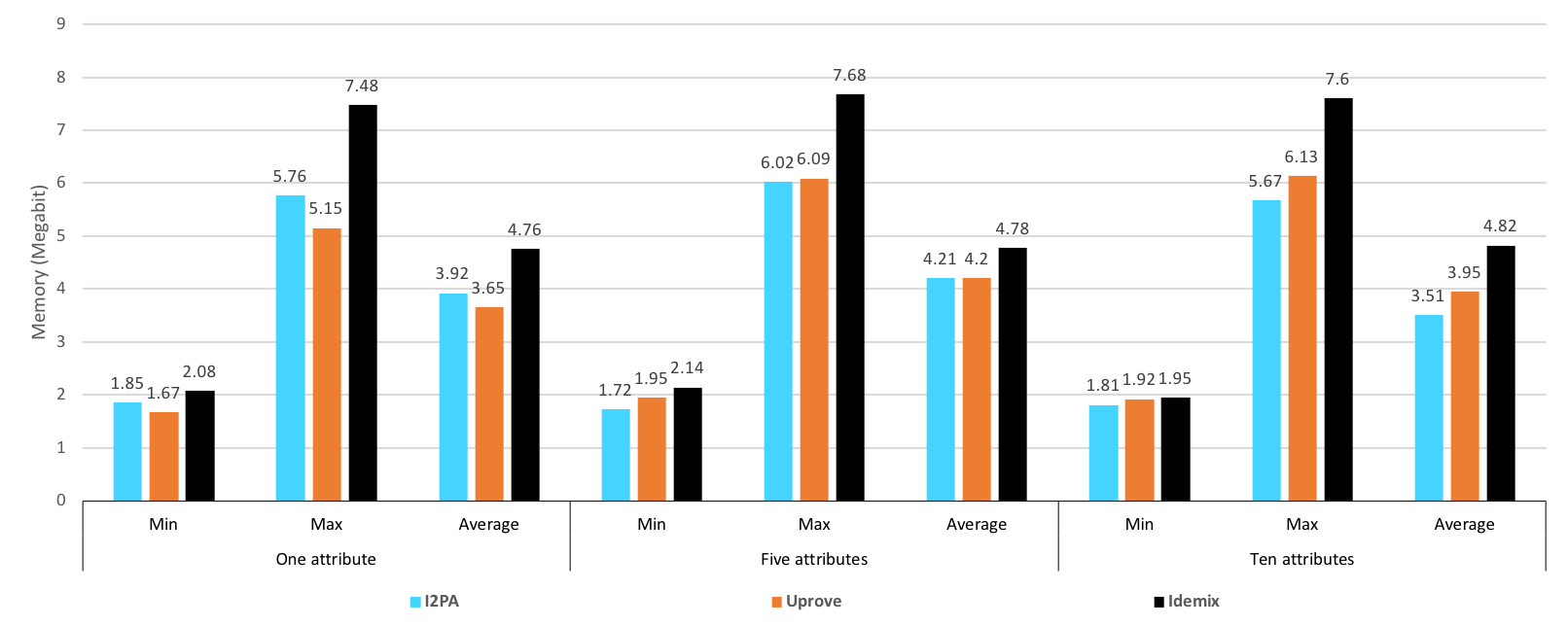} 
 \caption{ Issuance memory usage at issuer side }
 \label{fig:memory-issuer}
}
\end{figure}

\textbf{At issuer side: }
\label{label-issuing-issuer}
Recorded results  from issuance of credentials made up of 1, 5, and 10 attributes respectively are presented in "Figure \ref{fig:memory-issuer}".

As shown in "Figure \ref{fig:memory-issuer}", the three schemes present nearby performance regarding memory usage at issuer side while issuing credentials. At first glance, this may seem paradoxical seeing keys' size (160 for \itwopa and \uprovev 1024 for Idemix). However, this can be explained by the usage of extended homogeneous coordinates while instantiating \uprove and \itwopap Nevertheless, in all cases, \idemix requires more resource in average and  records highest maxima. \uprove and \itwopav in three cases, have nearby average consumptions; 3.92 against 3.65 (respectively 4.21 against 4.2, and 3.51 against 3.95) for issuance of 1 attribute (respectively 5 and 10 attributes)

\begin{figure}[H]
\centering{
\includegraphics[	]{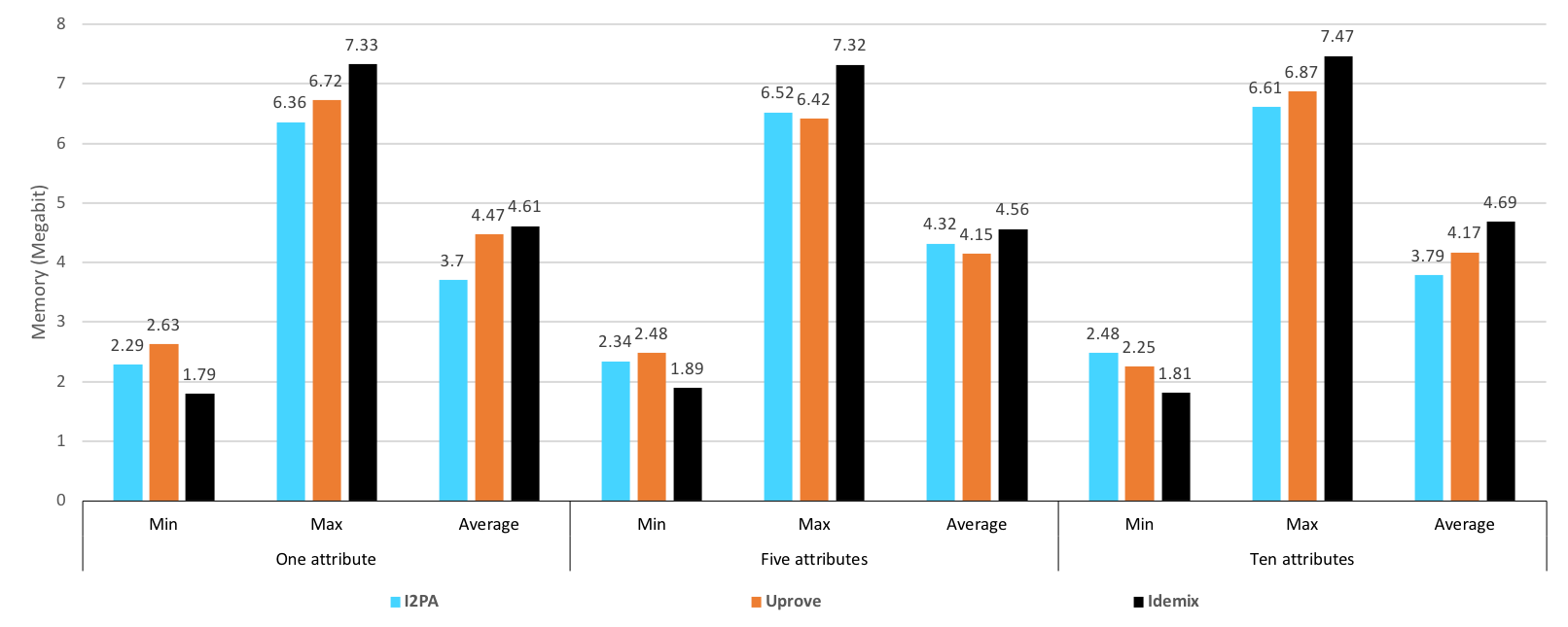} 
\caption{ Issuance memory usage at user side }
\label{fig:memory-user}
}
\end{figure}

\textbf{At user side: }
\label{label-verify-user}
This section describes memory usage at user side while issuing credentials of 1, 5, and 10 attributes respectively. We shall not evaluate the memory usage in the verification phase since the user only presents her credential; she performs no operation. Due to limitations noted in the mobile application while recording memory usage at user side, we recorded these results with a user implemented using Java socket and running in a Raspberry PI. The "Figure \ref{fig:memory-user}" is an illustration of recorded results.

"Figure \ref{fig:memory-user}" shows that, as we have already pointed out in the limitations  (Section \ref{label-limitations}), \idemix has very low minima (1.79, 1.89, and 1.81) compared to other schemes (2.29, 2.34, and 2.48 for \itwopav 2.63, 2.48, and 2.25 for U-prove). Despite the fact that three schemes present nearby consumptions, \idemix has higher maxima and requires more resources on average.

\subsubsection{Verification}

\begin{figure}[H]
 \centering{
 \includegraphics[width=0.9\linewidth]{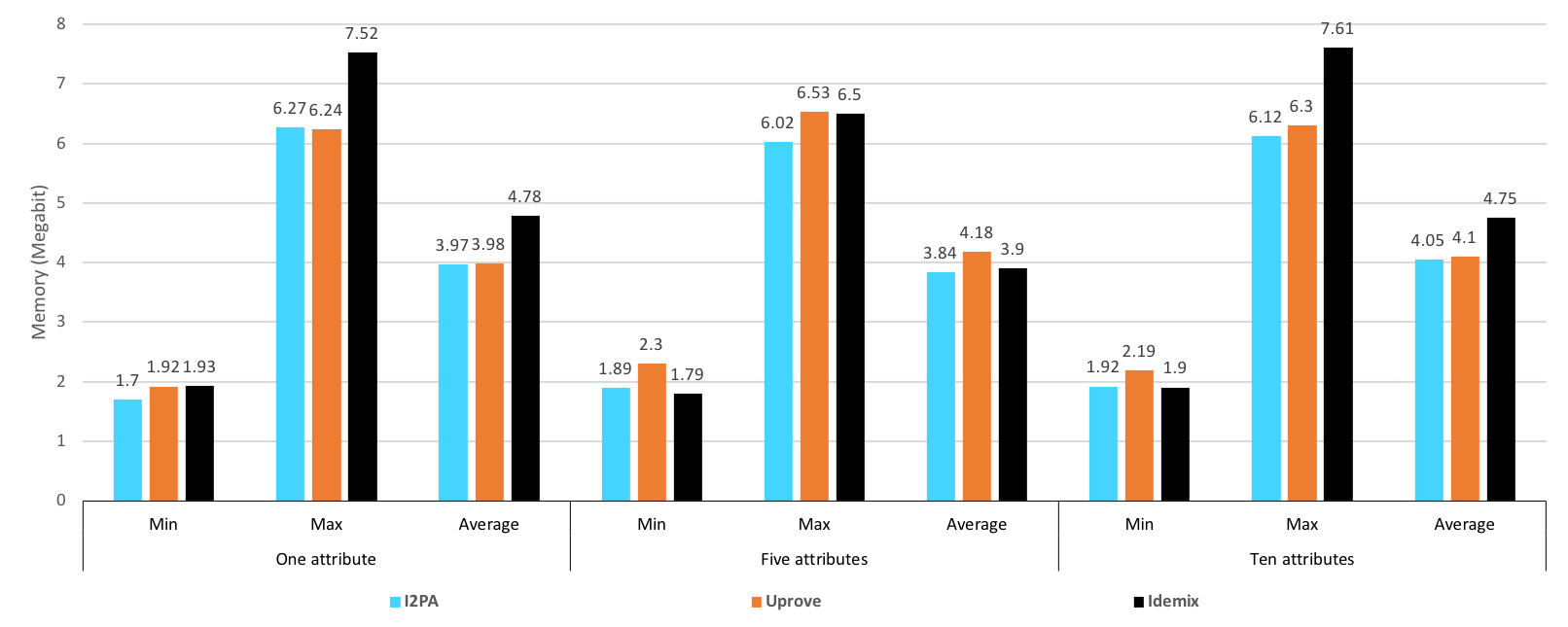}
 \caption{ Verification memory usage at verifier side }
 \label{fig:memory-verifier}
}
\end{figure}

If the credentials to verify are generated and stored beforehand, this may greatly affect memory usage and then influences recorded results. We verify a credential after generating it. This frees the memory once the credential is verified. "Figure \ref{fig:memory-verifier}" illustrates recorded results while verifying credentials of 1, 5, and 10 attributes respectively.

"Figure \ref{fig:memory-verifier}" shows that, globally, tendencies recorded here do not differ from those presented in  previous sections. We can note that on average, \idemixv requires more memory than \itwopa and \uprovep Except for the verification of credentials made up of 5 attributes, \idemix presents the highest average values. \itwopav in all three cases, has an average value smaller than that present by \uprove and \idemixp

%-----------------------
%TIME 
%-----------------------
\subsection{Time Evaluation}
\label{time-comparison}

This section describes results recorded regarding computing time. These results concern 100 simulations involving credentials of 1, 5, and 10 attributes respectively. We shall consider the time required at user side to get a credential from an issuer as well as the one required to have a credential verified by a verifier.

%-----------------------
%ISSUANCE
%-----------------------
\subsubsection{Issuance}
\label{label-issuing-issuer-one}

\begin{figure}[H]
 \centering{
 \includegraphics[width=0.9\linewidth]{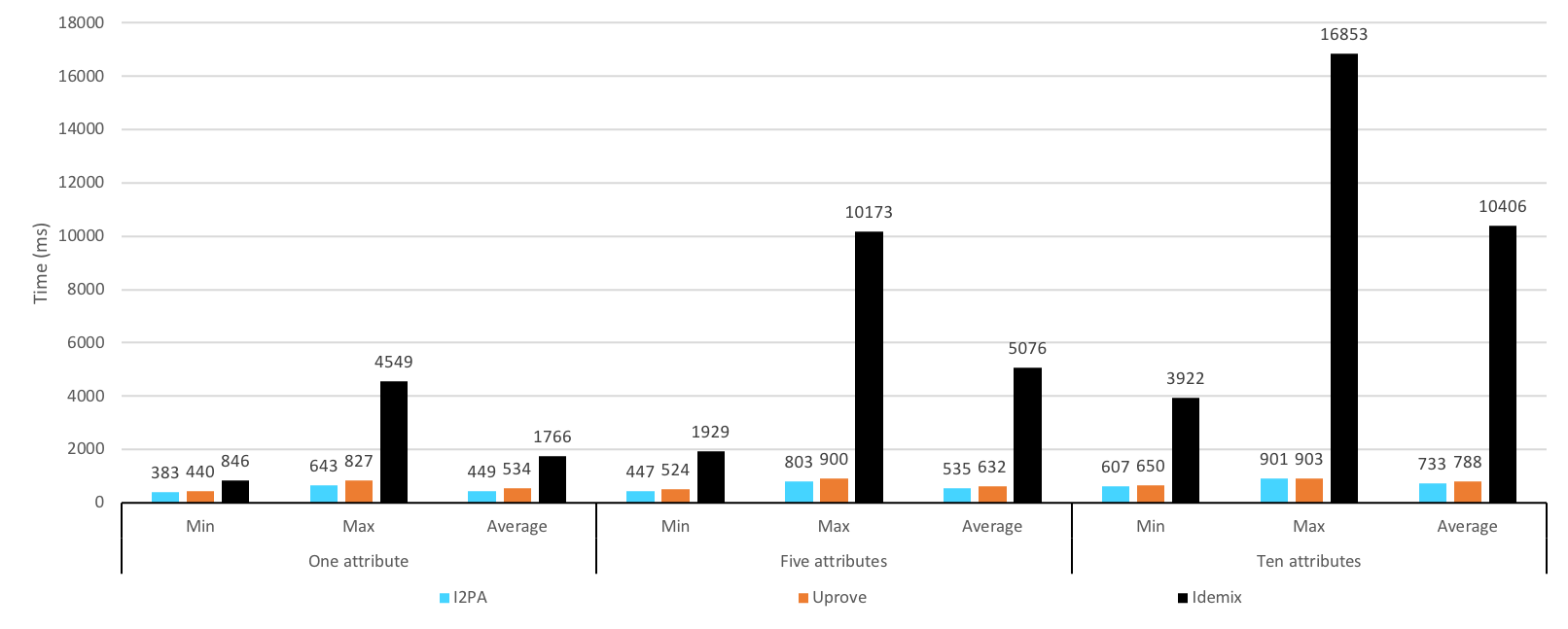}
 \caption{ Time issuance comparison }
 \label{fig:issue-time}
}
\end{figure}

Results recorded from issuance of credentials made up of 1, 5, and 10 attributes are illustrated in "Figure \ref{fig:issue-time}". We illustrate minima, maxima, as well as average values.

"Figure \ref{fig:issue-time}" shows that \itwopa and \uprove have similar performance regarding computing time efficiency. However, \itwopa presents more interesting result than \uprovep \idemixv meanwhile, has very low performance compared to \itwopa and \uprovep The time it requires for issuance is on average at least 3 times (respectively 8 and 13) more important than that required by \itwopa and \uprove for issuance of credential made up of 1 attribute (respectively 5 and 10 attributes). Regarding distribution of time for credential containing 1 attribute (respectively 5 and 10 attributes), 33\% (respectively 45\% and 50\%) of simulations have duration higher or equal to the average for \itwopav 38\% (respectively 45\% and 49\%) for \uprovev and 35\%  (respectively 39\% and 51\%) for \idemixp Finally, regarding computing time efficiency, \itwopa and \uprove present more interesting result than \idemixp What should be the number of attributes (1, 5, or 10), \itwopa and \uprove require less than 1 second for credential issuance, what can be considered as relevant.

%-----------------------
%ISSUANCE
%-----------------------
\subsubsection{Verification:}
\label{time-verification}

\begin{figure}[H]
 \centering{
 \includegraphics[width=0.9\linewidth]{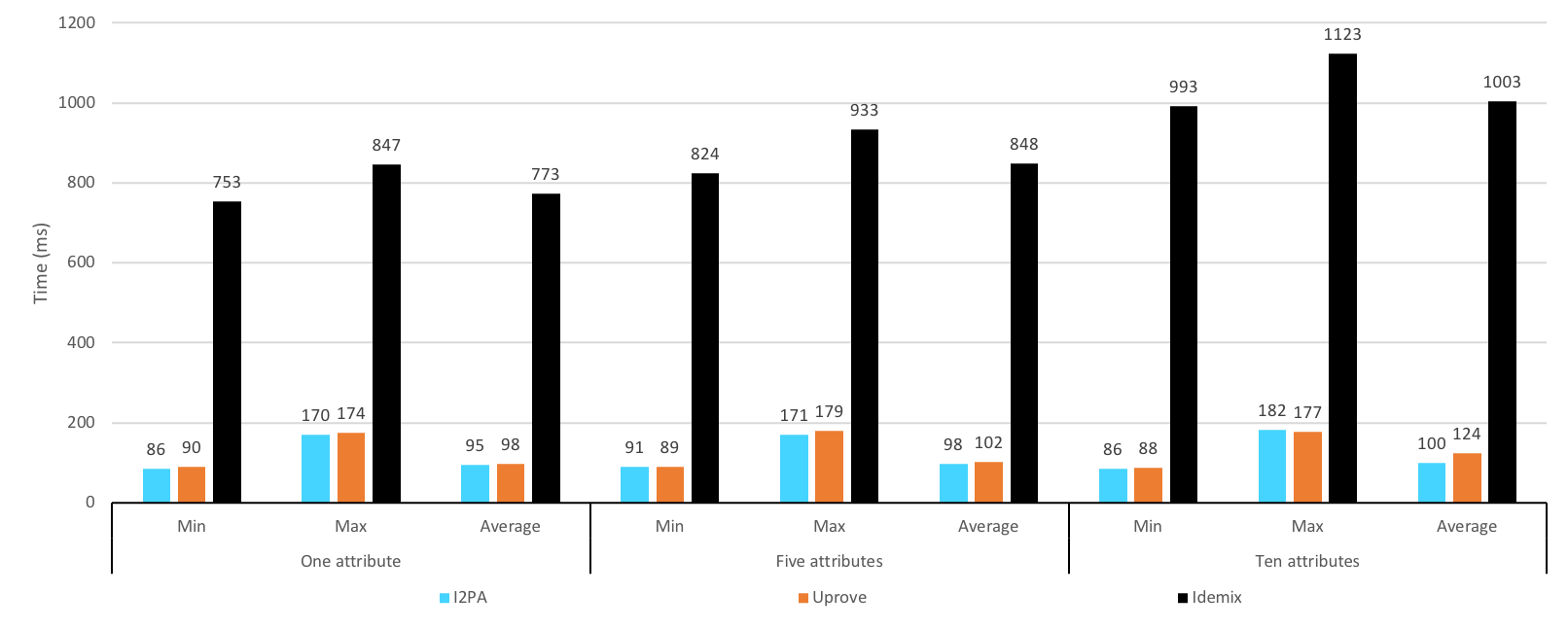}
 \caption{ Time verification comparison }
 \label{fig:verify-time}
}
\end{figure}

As for the issuance, this section describes recorded results while verifying 100 credentials of 1, 5, and 10 attributes respectively. "Figure \ref{fig:verify-time}" illustrates registered results.

"Figure \ref{fig:verify-time}" shows that, as for the issuance, \itwopa and \uprove present similar performances on the verification protocol regarding computing time efficiency. However, except the maximum recorded during the verification  of credentials made up of 10 attributes, \itwopa requires less computing time compared to \uprovep \idemixv meanwhile, has very low performance compared to \itwopa and \uprovep The time it requires for verification is on average at least 7 times (respectively 8 and 8) more important than that required by \itwopa and \uprove for verification of credentials made up of 1 attribute (respectively 5 and 10 attributes). Regarding distribution of time for credentials of 1 attribute (respectively 5 and 10 attributes), 22\% (respectively 10\% and 21\%) of simulations have duration higher or equal to the average for \itwopav 13\% (respectively 13\% and 50\%) for \uprove and 39\%  (respectively 54\% and 12\%) for \idemix. Finally, regarding computing time efficiency, we can safely assert that \itwopa and \uprove present more interesting result than \idemix on verification protocol. They can thus be envisaged in a context of resource-constrained devices.

%1: (I2PA,U-prove,Idemix)= (22,13,39)
%5: (I2PA,U-prove,Idemix)= (10,13,54)
%10: (I2PA,U-prove,Idemix)= (21,50,12)

%-----------------------
%CONCLUSION AND FUTURE WORKS
%-----------------------
\section{Conclusion and Future Works}
\label{label-conclusion-future-works}
\label{label-conclusion}

In this paper, the performance evaluation of  \itwopav \uprovev and \idemix we conducted  in low-resource devices, was focused in evaluating computing time and memory usage efficiency. Three types of conclusions can be drawn:

\begin{itemize}
\item {
In terms of memory usage at issuer, user or verifier sides, \itwopav \uprovev and \idemix present nearby consumptions if \itwopa and \uprove are instantiated using ECC and ECH representation. However, in average, \idemix requires more memory than \itwopa and \uprovep
}
\item {
In terms of computing time efficiency, \idemix has very low performances compared to \itwopa and \uprovep The time it requires for issuance (respectively verification) is at least 3 times (respectively 7 times) more important than that requires by \itwopa and \uprovep
}
\item {
Even though EHC representation speeds up operations over the curve, it increases memory usage.
}
\end{itemize}

Finally, for computing time and  memory usage efficiency criteria, \itwopa and \uprove are two schemes that can be envisaged in an IoT context. However, for an effective choice, other criteria must be taken into account including issuance unlinkability, multi-show unlinkability, selective disclosure, randomization, etc. This evaluation, for reasons of completeness,  could be extended by studying randomization and selective disclosure protocols efficiency as well as bandwidth usage.

\bibliographystyle{IEEEtran}
\bibliography{IEEEabrv,IEEEexample}

\end{document}